\documentstyle[12pt]{article}
\newcommand{\eq}{\begin{equation}}
\newcommand{\en}{\end{equation}}
\newcommand{\eqn}{\begin{eqnarray}}
\newcommand{\enn}{\end{eqnarray}}
\newcommand{\CR}{\nonumber \\}
\newcommand{\pa}{\partial}

\newcommand{\D}{\delta}          

\newcommand{\E}{\epsilon}
\newcommand{\La}{\Lambda}
\newcommand{\p}{\Phi}
\newcommand{\bra}{\langle}
\newcommand{\ket}{\rangle}
\newcommand{\lm}{\lambda}
\newcommand{\s}{\sigma}
\begin{document}

\begin{titlepage}
\null
\begin{flushright} 
hep-th/9607151  \\
UTHEP-340 \\
July, 1996
\end{flushright}
\vspace{0.5cm} 
\begin{center}
{\Large \bf 
Confining Phase of N=1 Supersymmetric 

Gauge Theories and N=2 Massless Solitons
\par}
\lineskip .75em
\vskip2.5cm
\normalsize
{\large Seiji Terashima} \quad and \quad {\large Sung-Kil Yang}
\vskip 1.5em
{\it Institute of Physics,  University of Tsukuba,  Ibaraki 305,  Japan}
\vskip3cm
{\bf Abstract}
\end{center} \par
Effective superpotentials for the phase with a confined photon are obtained
in $N=1$ supersymmetric gauge theories. We use the results to derive the
hyperelliptic curves which describe the Coulomb phase of $N=2$ theories with
classical gauge groups, and thus extending the prior result for $SU(N_c)$
gauge theory by Elitzur et al. Moreover, adjusting the coupling constants in
$N=1$ effective superpotentials to the values of $N=2$ non-trivial critical
points we find new classes of $N=1$ superconformal field theories with an
adjoint matter with a superpotential.

\end{titlepage}
\renewcommand{\thefootnote}{\arabic{footnote}}
\setcounter{footnote}{0}
\baselineskip=0.7cm

%\section{Introduction}
 
Exact descriptions of strong-coupling dynamics of supersymmetric gauge theories
in four dimensions have been obtained on the basis of the idea of 
duality and holomorphy 
\cite{Se}, \cite{SeWi}, \cite{SeWi2}, \cite{Se2}.
In these exact solutions an important feature in common is that 
singularities of quantum moduli space of the theory correspond to 
the appearance of massless solitons.
Near the singularity, therefore, we observe interesting non-perturbative 
properties of supersymmetric gauge theories.

In an $N=2$ case the Coulomb phase admits
a beautiful mathematical description according to which
massless solitons are recognized as vanishing cycles 
associated with the degeneracy of hyperelliptic curves.
In order to explore physics near $N=2$ singularities 
the microscopic superpotential explicitly breaking $N=2$ to $N=1$ 
supersymmetry is often considered 
\cite{SeWi}, \cite{SeWi2}, \cite{InSe}, \cite{ArDo}, \cite{DPK}. 
Examining the resulting superpotential 
for a low-energy effective Abelian theory it is found that 
the generic $N=2$ vacuum is lifted and only the singular loci
of moduli space remain as the $N=1$ vacua where monopoles or dyons 
can condense.

Alternatively we may start with a microscopic $N=1$ theory which 
we introduce by perturbing an $N=2$ theory by adding a tree-level
superpotential built out of the Casimirs of the adjoint field
in the vector multiplet \cite{InSe}, \cite{ElFoGiRa}, \cite{ElFoGiRa2}.
Let us concentrate on a phase with a single confined
photon in our $N=1$ theory which 
corresponds to the classical $SU(2) \times U(1)^{r-1}$
vacua with $r$ being the rank of the gauge group.
Then the low-energy effective theory
containing non-perturbative effect provides us with the
data of the vacua with massless solitons \cite{In}, \cite{InSe}.
This will enable us to reconstruct hyperelliptic curves for $N=2$ theories.

Along this line of thought Elitzur et al. \cite{ElFoGiInRa}
have recently obtained 
$N=1$ effective superpotentials from which 
the curves for the Coulomb phase of $N=2$ $SU(N_c)$ gauge theories
can be verified. Our purpose in this paper is to extend
their result to the case of arbitrary classical gauge group.
In what follows we first derive $N=1$ effective superpotentials
which can be used to reproduce the $N=2$ curves for classical gauge groups.
We next discuss $N=1$ superconformal field theories based on our results 
for superpotentials.

%%%%%%%%%%%%%%%%%%%%%%%%%%%
%\section{$SU(N_c)$ case}

We begin with briefly reviewing the results of \cite{ElFoGiInRa} 
on $SU(N_c)$ gauge theories.
The gauge symmetry breaks down to $U(1)^{N_c-1}$ in the 
Coulomb phase of $N=2$ $SU(N_c)$ Yang-Mills theories.
Near the singularity of a single massless dyon 
we have a photon coupled to the light dyon hypermultiplet
while the photons for the rest $U(1)^{N_c-2}$ factors remain free.
We now perturb the theory by adding a tree-level superpotential
\eq
W=\sum_{n=1}^{N_c} g_n u_n, 
\hskip10mm  u_n=\frac{1}{n} {\rm Tr}\,  \p^n ,
\label{r1}
\en
where $\p$ is the adjoint $N=1$ superfield in the $N=2$ vectormultiplet and
$g_1$ is an auxiliary field implementing ${\rm Tr}\, \p =0$.
In view of the macroscopic theory, we see that under the perturbation 
by (\ref{r1}) only the $N=2$ singular loci survive as the 
$N=1$ vacua where a single photon is confined
and the $U(1)^{N_c-2}$ factors decouple.

The result should be directly recovered when we start with the 
microscopic $N=1$ $SU(N_c)$ 
gauge theory which is obtained from $N=2$ $SU(N_c)$ Yang-Mills
theory perturbed by (\ref{r1}).
For this we study the vacuum with unbroken $SU(2) \times U(1)^{N_c-2}$.
The classical vacua of the theory are determined by the equation of motion
$W'(\p)=\sum_{i=1}^{N_c} g_i \p^{i-1}=0$. Then the roots $a_i$ of 
\eq
W'(x)=\sum_{i=1}^{N_c} g_i x^{i-1}=
g_{N_c} \prod_{i=1}^{N_c-1} (x-a_i)
\label{suneigen}
\en
give the eigenvalues of $\p$. In particular the unbroken 
$SU(2) \times U(1)^{N_c-2}$ vacuum is described by 
\eq
\Phi ={\rm diag} (a_1, a_1, a_2, a_3, \cdots , a_{N_c-1}).
\en
In the low-energy limit the adjoint superfield for $SU(2)$ becomes massive and
will be decoupled. We are then left with an $N=1$ $SU(2)$ Yang-Mills theory 
which is in the confining phase and the
photon multiplets for $U(1)^{N_c-2}$ are decoupled.

The relation between the high-energy $SU(N_c)$ scale $\La$ and the low-energy 
$SU(2)$ scale $\La_L$ is determined by first matching at the
scale of $SU(N_c)/SU(2)$ $W$ bosons and then by matching at the $SU(2)$
adjoint mass $M_{\rm ad}$. One finds \cite{KuScSe}, \cite{ElFoGiInRa}
\eq
\La^{2 N_c}=
{\La_L}^{3 \cdot 2} 
\left ( \prod_{i=2}^{N_c-1} (a_1-a_i) \right )^2
(M_{\rm ad})^{-2}.
\label{matching}
\en
To compute $M_{\rm ad}$ we decompose 
\eq
\Phi=  \Phi_{cl} + \delta \Phi + \delta \tilde{\Phi},
\label{decompose}
\en
where 
$ \delta \Phi $ denotes the fluctuation along the unbroken $SU(2)$ 
direction and $ \delta \tilde{\Phi}$ along the other directions.
Substituting this into $W$ we have
\eqn
W & = & 
W_{cl} + \sum_{i=2}^{N_c} g_i \frac{i-1}{2}
\; {\rm Tr}\, ( \delta \Phi^2  \Phi_{cl}^{i-2} )+\cdots \CR
& =& W_{cl} + \frac{1}{2} W''(a_1) \; {\rm Tr}\, \D \Phi^2+\cdots \CR
& =& W_{cl} + \frac{1}{2} \; g_{N_c} \prod^{N_c}_{i=2} (a_1-a_i)\;
{\rm Tr}\, \D \Phi^2 +\cdots ,
\enn
where $[ \D \Phi,  \Phi_{cl} ]=0$ has been used and $W_{cl}$ is the 
tree-level superpotential evaluated in the classical vacuum.
Hence, $M_{\rm ad}=g_{N_c} \prod^{N_c-1}_{i=2} (a_1-a_i)$ and the relation
(\ref{matching}) reduces to
\eq
{\La_L}^6 = g_{N_c}^2 \La^{2 N_c}.
\en
Since the gaugino condensation dynamically generates the 
superpotential in the $N=1$ $SU(2)$ theory
the low-energy effective superpotential finally takes the form 
\cite{ElFoGiInRa}
\eq
W_L= W_{cl} \pm  2 {\La_L}^3 = W_{cl} \pm 2 g_{N_c} \La^{N_c}.
\label{wl}
\en

We simply assume here that the superpotential (\ref{wl}) is exact for any 
values of the parameters. (This is equivalent to assume 
$W_{\Delta}=0$ \cite{In}, \cite{ElFoGiInRa}.) From (\ref{wl}) we obtain
\eq
\langle u_n \rangle = {\partial W_L \over \partial g_n}
=u_n^{cl}(g) \pm 2 \La^{N_c} \delta_{n, N_c} 
\label{qvacua}
\en
with $u_n^{cl}$ being a classical value of $u_n$. As we argued above these 
vacua should correspond to the singular loci of $N=2$ massless dyons. This
can be easily confirmed by plugging (\ref{qvacua}) in the $N=2$ $SU(N_c)$
curve \cite{KlLeYaTh}, \cite{ArFa}
\eq
y^2 = {\left \langle {\rm det} (x-\Phi) \right \rangle}^2- 4 \La^{2 N_c} 
= \left( x^{N_c}- \sum_{i=2}^{N_c} \bra s_i \ket x^{N_c-i} \right)^2
- 4 \La^{2 N_c},
\en
where
\eq
ks_k+\sum_{i=1}^k i s_{k-i} u_i=0, \hskip10mm k=1,2,\cdots
\en
with $s_0=-1$ and $s_1=u_1=0$. We have
\eqn
y^2 &=& \left ( x^{N_c}- s_2^{cl} x^{N_c-2}- \cdots -s_{N_c}^{cl} \right ) 
\left ( x^{N_c}- s_2^{cl} x^{N_c-2}- \cdots 
-s_{N_c}^{cl} \pm 4 \La^{N_c} \right ) \CR
 &=& (x-a_1)^2 (x-a_2) \cdots (x-a_{N_c-1}) \, 
\Big( (x-a_1)^2 (x-a_2) \cdots (x-a_{N_c-1}) \pm 4 \La^{N_c} \Big).
\label{su}
\enn
Since the curve exhibits the quadratic degeneracy we are exactly at the
singular point of a massless dyon in the $N=2$ $SU(N_c)$ Yang-Mills vacuum.

%%%%%%%%%%%%%%%%%%%%%%%%%%%%%%
%\section{$SO(2 N_c)$ case}

Let us now apply our procedure to the $N=2$ $SO(2 N_c)$ Yang-Mills theory.
We take a tree-level superpotential to break $N=2$ to $N=1$ as 
\eq
W=\sum_{n=1}^{N_c-1} g_{2 n} u_{2 n} + \lm v,
\label{soeventree}
\en
where 
\eqn
&& u_{2 n} =\frac{1}{2 n} {\rm Tr}\, \Phi^{2 n}, \CR
&& v ={\rm Pf}\, \Phi=\frac{1}{2^{N_c} N_c !} \E_{i_1 i_2 j_1 j_2 \cdots}
\Phi^{i_1 i_2} \Phi^{j_1 j_2} \cdots 
\enn
and the adjoint superfield $\Phi$ is an antisymmetric 
$2 N_c \times 2 N_c$ tensor.
This theory has classical vacua which satisfy the condition
\eq
W'(\Phi)=\sum_{i=1}^{N_c-1} g_{2 i} (\Phi^{2 i-1})_{ij}-
\frac{\lm}{2^{N_c} (N_c-1) !} \E_{\: i\: j \: k_1 k_2 l_1 l_2 \cdots}
\Phi^{k_1 k_2} \Phi^{l_1 l_2} \cdots=0.
\label{wd}
\en
For the skew-diagonal form of $\p$
\eq \p={\rm diag}
(\s_{2}e_0,\; \s_{2}e_1,\; \s_{2}e_2, \cdots, \s_{2}e_{N_c-1}), \hskip10mm
\s_{2}=i \left( \begin{array}{cc} 0 &  -1 \\ 1 & 0 \end{array} \right )
\en
the vacuum condition (\ref{wd}) becomes
\eq
\sum_{i=1}^{N_c-1} g_{2 i} (-1)^{i-1} {e_n}^{2 i-1}+ (-i)^{N_c}
\frac{\lm}{2 e_n} \prod_{i=0}^{N_c-1} e_i=0 \;, \hskip10mm 0 \leq n \leq N_c-1.
\label{ai}
\en
Thus we see that $e_n \, (\not= 0)$ are the roots of $f(x)$ defined by 
\eq
f(x)=\sum_{i=1}^{N_c-1} g_{2 i} x^{2 i}+d,
\label{fpoly}
\en
where we put $d=(-i)^{N_c} \frac{1}{2} \lm \prod_{i=0}^{N_c-1} e_i $.

Since our main concern is the vacuum with a single confined photon 
we focus on the unbroken $SU(2) \times U(1)^{N_c-1}$ vacuum. Thus writing  
(\ref{fpoly}) as
\eq
f(x)=g_{2(N_c-1)} \prod_{i=1}^{N_c-1} (x^2-a_i^2),
\en
we take
\eq
\Phi  =  
{\rm diag} (\s_{2}a_1,\; \s_{2}a_1,\;  \s_{2}a_2, \cdots, \s_{2}a_{N_c-1})
\en 
with $d = (-i)^{N_c} \frac{1}{2} \lm a_1^2 \prod_{i=2}^{N_c-1} a_i$.
We then make the scale matching between the 
high-energy $SO(2 N_c)$ scale $\La$ and the low-energy $SU(2)$ scale 
$\La_L$. Following the steps as in the $SU(N_c)$ case yields
\eq
\La^{2 \cdot 2( N_c-1)}={\La_L}^{3 \cdot 2}
\left ( \prod_{i=2}^{N_c-1} (a_1^2-a_i^2) \right )^2
(M_{\rm ad})^{-2}, 
\label{so2nmatching}
\en
where the factor arising through the Higgs mechanism is easily calculated 
in an explicit basis of $SO(2 N_c)$. In order to evaluate the $SU(2)$ adjoint 
mass $M_{\rm ad}$ we first substitute the decomposition (\ref{decompose})
in $W$ and proceed as follows:
\eqn
W
& = & W_{cl} + \sum_{i=1}^{N_c-1} g_i \frac{2 i-1}{2}
\; {\rm Tr}\, ( \delta \Phi^2  \Phi_{cl}^{2 i-2} )
+ \lm \left ( {\rm Pf}_4 \D \Phi \right ) 
\left({\rm Pf}_{2(N_c-2)} \Phi_{cl} \right ) +\cdots \CR
& = & W_{cl} + \sum_{i=1}^{N_c-1} g_i \frac{2 i-1}{2}
\; {\rm Tr}\, ( \delta \Phi^2  \Phi_{cl}^{2 i-2} )
+ \lm \left ( \frac{1}{4} {\rm Tr}\, \D \Phi^2 \right ) 
\left ( \prod_{k=2}^{N_c-1} (-i a_k) \right ) +\cdots \CR
& =& W_{cl} + \left. \frac{1}{2} \frac{{\rm d}}{{\rm d} x} 
\left( \frac{f(x)}{x} \right ) \right |_{x=a_1} 
{\rm Tr}\, \D \Phi^2 +\cdots \CR
& =& W_{cl} +  g_{2(N_c-1)} \prod^{N_c-1}_{i=2} (a_1^2-a_i^2)
\, {\rm Tr}\, \D \Phi^2  +\cdots, 
\enn
where ${\rm Pf}_4$ is the Pfaffian of a upper-left $4 \times 4$ sub-matrix and
${\rm Pf}_{2(N_c-2)} $ is the Pfaffian of a lower-right 
$ 2(N_c-2)\times 2(N_c-2)$ sub-matrix. Thus we observe that $M_{\rm ad}$
cancels out the Higgs factor in (\ref{so2nmatching}), which leads to
${\La_L}^6 =  g_{2(N_c-1)}^2 \La^{4(N_c-1)}$. The low-energy 
superpotential is now given by 
\eq
W_L= W_{cl} \pm 2 {\La_L}^3 = W_{cl} \pm 2 g_{2(N_c-1)} \La^{2(N_c-1)},
\en
where the second term is due to the gaugino condensation in the low-energy
$SU(2)$ theory. 

The vacuum expectation values of gauge invariants are obtained from $W_L$ as
\eqn
&& \langle u_{2 n} \rangle = {\pa W_L \over \pa g_{2n}}=
u_{2 n}^{cl}(g,\lambda) \pm 2 \La^{2(N_c-1)} \delta_{n, N_c-1}, \CR
&& \langle v \rangle ={\pa W_L \over \pa \lm }=v_{cl}(g,\lambda).
\label{so2nsingular}
\enn
The curve for  $N=2\;SO(2 N_c)$ is known to be \cite{BrLa}
\eqn
y^2 &=& {\left \langle {\rm det} (x-\Phi) \right \rangle}^2
-4 \La^{4( N_c-1)} x^4  \CR
&=& \Big( x^{2N_c}- \sum_{i=1}^{N_c-1} \bra s_{2i} \ket x^{2(N_c-i)}
+\bra v \ket^2 \Big)^2-4 \La^{4( N_c-1)} x^4,
\enn
where
\eq
ks_k+\sum_{i=1}^k i s_{k-i} u_{2i}=0, \hskip10mm k=1,2,\cdots
\label{defs}
\en
with $s_0=-1$. At the values (\ref{so2nsingular}) of the moduli coordinates
we see the quadratic degeneracy 
\eqn
y^2 &=& \left ( x^{2 N_c}-   s_2^{cl}   x^{2 (N_c-1)}- \cdots 
-  s^{cl}_{2 (N_c-1)} x^2 +v_{cl}^2   \right ) \CR 
 & & \;\; \times 
\left ( x^{2 N_c}-   s_2^{cl}   x^{2 (N_c-1)}- \cdots 
-  s_{2 (N_c-1)}^{cl} x^2  +v_{cl}^2  \pm 4 \La^{2 (N_c-1)} x^2 \right ) \CR
 &=& (x^2-a_1^2)^2 (x^2-a_2^2) \cdots (x^2-a_{N_c-1}^2) \CR
& & \;\; \times
\Big((x^2-a_1^2)^2 (x^2-a_2^2) \cdots (x^2-a_{N_c-1}^2) 
\pm 4 \La^{2 (N_c-1)} x^2 \Big).
\enn
This is our desired result. Notice that the apparent singularity at 
$\langle v \rangle=0$ is not realized in our $N=1$ theory. Thus the point
$\langle v \rangle=0$ does not correspond to massless solitons in agreement
with the result of \cite{BrLa}.

%%%%%%%%%%%%%%%%%%%%%%%%%%%%%%%
%\section{$SO(2 N_c+1)$ case}

Our next task is to study the $SO(2 N_c+1)$ gauge theory.
A tree-level superpotential breaking $N=2$ to $N=1$ is assumed to be 
\eq
W=\sum_{n=1}^{N_c} g_{2 n} u_{2 n}, \hskip10mm 
u_{2 n} = \frac{1}{2 n} {\rm Tr}\, \Phi^{2 n}.
\label{sooddtree}
\en
The classical vacua obey
$W'(\Phi)=\sum_{i=1}^{N_c} g_{2 i} \Phi^{2 i-1}=0$.
The eigenvalues of $\p$ are given by the roots $a_i$ of
\eq
W'(x)=\sum_{i=1}^{N_c} g_{2 i} x^{2 i-1}=
g_{2 N_c} x \prod_{i=1}^{N_c-1} (x^2-a_i^2).
\label{sooddmotion}
\en
As in the previous consideration we take 
the $SU(2) \times U(1)^{N_c-1}$ vacuum.
Notice that there are two ways of  breaking  $SO(2 N_c+1)$ to 
$SU(2) \times U(1)^{N_c-1}$. One is to take all the eigenvalues distinct
(corresponding to $SO(3) \times U(1)^{N_c-1}$). The other is to choose
two eigenvalues coinciding and the rest distinct
(corresponding to $SU(2) \times U(1)^{N_c-1}$ with $a_i \not= 0$).
Here we examine the latter case
\eq
\Phi=
{\rm diag}(\s_{2}a_1,\; \s_{2}a_1,\; \s_{2}a_2, \cdots, \s_{2}a_{N_c-1},\; 0),
\hskip10mm 
\s_{2}=i \left( \begin{array}{cc} 0 &  -1 \\ 1 & 0 \end{array} \right ).
\en
In this vacuum the high-energy $SO(2 N_c+1)$ scale $\La$ and the low-energy
 $SU(2)$ scale $\La_L$ are related by
\eq
\La^{2 \cdot (2 N_c-1)}={\La_L}^{3 \cdot 2}
a_1^2 \left ( \prod_{i=2}^{N_c-1} (a_1^2-a_i^2) \right )^2
(M_{\rm ad})^{-2},
\en
where the $SU(2)$ adjoint mass $M_{\rm ad}$ is read off from
\eqn
W
& = & W_{cl} + \sum_{i=1}^{N_c} g_{2 i} \frac{2 i-1}{2}
\; {\rm Tr}\, ( \delta \Phi^2  \Phi_{cl}^{2 i-2} )+\cdots \CR
& =& W_{cl} + \frac{1}{2} W''(a_1) \; {\rm Tr}\, \D \Phi^2+\cdots \CR
& =& W_{cl}+g_{2 N_c} a_1^2 \prod^{N_c-1}_{i=2} (a_1^2-a_i^2)\,
{\rm Tr}\, \D \Phi^2  +\cdots.
\enn
So, we obtain ${\La_L}^6 =  g_{2 N_c}^2 a_1^2 \La^{2(2 N_c-1)}$. The low-energy
effective superpotential becomes
\eq
W_L= W_{cl} \pm 2 {\La_L}^3 = 
W_{cl} \pm 2 g_{2 N_c} a_1 \La^{2 N_c-1}.
\label{wll}
\en
If we assume $W_{\Delta}=0$ the expectation values $\bra u_{2i}\ket$
are calculated from $W_L$ by expressing $a_1$ as a function of $g_{2 i}$.

For the sake of illustration let us discuss the $SO(5)$ theory 
explicitly. From (\ref{wll}) we get
\eqn
\langle u_{2} \rangle & = & 2 a_1^2 \pm \frac{1}{a_1} \La^3, \CR
\langle u_{4} \rangle & = &  a_1^4  \mp a_1 \La^3 
\label{r3}
\enn
and $a_1^2=-g_2/g_4$. We eliminate $a_1 $ from (\ref{r3}) to obtain 
\eq
27 \La^{12} - \La^6 u_2^3 + 36 \La^6 u_2 u_4 -u_2^4 u_4 
+8 u_2^2 u_4^2 -16 u_4^3=0.
\label{r4}
\en
This should be compared with the $N=2$ $SO(5)$ discriminant \cite{DaSu}
\eq
s_2^2 (27 \La^{12} - \La^6 s_1^3-36 \La^6 s_1 s_2 +s_1^4 s_2 
+8 s_1^2 s_2^2 +16 s_2^3)^2=0,
\label{r6}
\en
where $s_1=u_2$ and $s_2=u_4-u_2^2/2$ according to (\ref{defs}). Thus we
see the discrepancy between (\ref{r4}) and (\ref{r6}) which implies
that our simple assumption of $W_{\Delta} = 0$ does not work.
Inspecting (\ref{r4}) and (\ref{r6}), however, we notice how to remedy the
difficulty. Instead of (\ref{sooddtree}) we take a tree-level superpotential
\eq
W=g_{2} s_1+g_4 s_2=
g_2 u_2+g_4 \left(u_4-\frac{1}{2} u_2^2 \right).
\label{ws2}
\en
The classical vacuum condition is
\eq
 W'(\Phi)=(g_2 -g_4 u_2) \Phi +g_4 \Phi^3 =0.
\en
To proceed, therefore, we can make use of the results obtained in the
foregoing analysis just by making the replacement
\eqn
g_4 & \rightarrow & \tilde{g_4}=g_4, \CR
g_2 & \rightarrow & \tilde{g_2}=g_2-u_2 g_4.
\label{u2}
\enn
(especially evaluation of $M_{\rm ad}$ is not invalidated because
${\rm Tr}\, \D \Phi=0$.)
The eigenvalues of $\Phi$ are now determined in a self-consistent manner by
\eq
W'(x)=\tilde{g_2} x+\tilde{g_4} x^3=
\tilde{g_4} x \left( x^2+\frac{\tilde{g_2}}{\tilde{g_4}} \right)
=\tilde{g_4} x (x^2-a_1^2)=0.
\en
Then we have $u_2^{cl}=2 a_1^2=-2 \tilde{g_2}/\tilde{g_4}$ and
$\tilde{g_2} =-g_2$ from (\ref{u2}), which leads to
\eq
a_1^2=\frac{g_2}{g_4}.
\en
Substituting this in (\ref{wll}) we calculate $\bra s_i \ket$ and 
find the relation of $s_i$ which is precisely the discriminant (\ref{r6})
except for the classical singularity at $\bra s_2\ket =0$.

The above $SO(5)$ result indicates that an appropriate mixing term with
respect to $u_{2i}$ variables in a microscopic superpotential will be
required for $SO(2 N_c+1)$ theories. We are led to assume
\eq
W=\sum_{i=1}^{N_c-1} g_{2 i} u_{2 i} + g_{2 N_c} s_{N_c}
\label{ws}
\en
for the gauge group $SO(2 N_c+1)$ with $N_c \geq 3$. Then the following 
analysis is analogous to the $SO(5)$ theory. First of all notice that
\eq
s_{N_c}=u_{2 N_c} -u_{2(N_c-1)} u_2 +
(\hbox{polynomials of $u_{2 k}, \;1 \leq k < N_c-1$}).
\en
Therefore the eigenvalues of $\Phi$ are given by the roots of 
(\ref{sooddmotion}) with the replacement
\eqn
g_{2 N_c} & \rightarrow & \tilde g_{2 N_c}=g_{2 N_c}, \CR
g_{2 (N_c-1)} & \rightarrow & \tilde g_{2 (N_c-1)}=
g_{2 (N_c-1)} - u_2 g_{2 N_c}.
\enn
Then we have $u_2=a_1^2+\sum_{k=1}^{N_c-1} a_k^2=a_1^2-
 \tilde g_{2 (N_c-1)}/\tilde g_{2 N_c}$ and find
\eq
a_1^2=\frac{g_{2 (N_c-1)}}{g_{2 N_c}}.
\en
It follows that the effective superpotential is given by
\eq
W_L=W_L^{cl} \pm 2 \sqrt{g_{2 N_c} g_{2(N_c-1)}} \La^{2 N_c-1}.
\en

The vacuum expectation values of gauge invariants are obtained from $W_L$ as 
\eqn
\langle s_n \rangle & = &  s_n^{cl}(g), \hskip10mm  1 \leq n \leq N_c-2 \CR 
\langle s_{N_c-1} \rangle & = &  s_{N_c-1}^{cl}(g) 
\pm \frac{1}{a_1} \La^{2 N_c-1}, \CR
\langle s_{N_c} \rangle & = &  s_{N_c}^{cl}(g)
\pm a_1 \La^{2 N_c-1} .
\enn
For these $\bra s_i\ket$ the curve describing the $N=2\;SO(2 N_c+1)$ 
theory \cite{DaSu} is shown to be degenerate as follows:
\eqn
y^2 & = & {\left \langle {\rm det} (x-\Phi) \right \rangle}^2
- 4 x^2 \La^{2( 2 N_c-1)}  \CR
& = & (x^{2 N_c} - \langle s_1 \rangle x^{2(N_c-1)} -\cdots 
 - \langle s_{N_c-1} \rangle x^2- \langle s_{N_c} \rangle
+ 2 x \La^{2 N_c-1} ) \CR
&  & \; \times (x^{2 N_c} - \langle s_1 \rangle x^{2(N_c-1)} -
\cdots 
 - \langle s_{N_c-1} \rangle x^2- \langle s_{N_c} \rangle
- 2 x \La^{2 N_c-1} ) \CR
& = & \left\{ (x^2-a_1^2)^2 (x^2-a_2^2) \cdots (x^2-a_{N_c-1}^2) \pm
\La^{2 N_c-1} \left( -\frac{x^2}{a_1} -a_1 +2 x \right) \right\} 
\CR
&  & \; \times
\left\{ (x^2-a_1^2)^2 (x^2-a_2^2) \cdots (x^2-a_{N_c-1}^2) \pm
\La^{2 N_c-1} \left( -\frac{x^2}{a_1} -a_1 -2 x \right) \right\} 
\CR
& = & (x^2-a_1^2)^2 
\left( (x+a_1)^2 (x^2-a_2^2) \cdots (x^2-a_{N_c-1}^2) \mp 
\frac{\La^{2 N_c-1}}{a_1} \right) \CR
&  & \; \times
\left( (x-a_1)^2 (x^2-a_2^2) \cdots (x^2-a_{N_c-1}^2) \mp 
\frac{\La^{2 N_c-1}}{a_1} \right).
\enn
Thus we see the theory with the superpotential (\ref{ws}) recover the
$N=2$ curve correctly with the assumption $W_{\Delta}=0$. 
As in the $SO(2 N_c)$ case, 
the singularity at $\langle s_{ N_c} \rangle=0$, which corresponds to the
classical $SO(3) \times U(1)^{N_c-1}$ vacuum,
does not arise in our theory.

We remark that the particular form of superpotential (\ref{ws}) is not 
unique to derive the singularity manifold.
In fact we may start with a superpotential
\eq
W=\sum_{i=1}^{N_c-1} g_{2 i} \left(u_{2 i} +h_i(s) \right)+ 
g_{2 N_c} \left(s_{N_c}+h_{N_c}(s) \right), 
\label{r7}
\en
where $h_i(s)$ are arbitrary polynomials of $s_j$ with $j \geq N_c-2$,
to verify the $N=2$ curve. However, we are not allowed to take 
a superpotential such as $W=\sum_{i=1}^{N_c} g_{2 i} s_i$, 
because there are no $SU(2) \times U(1)^{N_c-1}$ vacua
(there exist no solutions for $\tilde g_{2( N_c-1)}$).
Note also that there are no $SO(3) \times U(1)^{N_c-1}$ vacua
in the theory with superpotential (\ref{r7}).

%%%%%%%%%%%%%%%%%%%%%%%%%%%%
%\section{$Sp(2 N_c)$ case}

Finally we discuss the $Sp(2 N_c)$ gauge theory.
The adjoint superfield $\Phi$ is a $2 N_c \times 2 N_c$ tensor which is
subject to
\eq
{}^t \Phi = J \Phi J \quad \Longleftrightarrow \quad J \Phi 
\;\; \hbox{is symmetric},
\en
where $J={\rm diag}(i\sigma_2, \cdots, i\sigma_2)$.
Let us assume a tree-level superpotential 
\eq
W=\sum_{n=1}^{N_c} g_{2 n} u_{2 n}, \hskip10mm
u_{2 n} = \frac{1}{2 n} {\rm Tr}\, \Phi^{2 n}.
\label{sptree}
\en
Then our analysis will become quite similar to that for $SO(2 N_c+1)$. The
classical vacuum with unbroken $SU(2) \times U(1)^{N_c-1}$ gauge group
corresponds to
\eq
J \Phi=
{\rm diag}(\s_{1}a_1,\; \s_{1}a_1,\; \s_{1}a_2,\cdots,\s_{1}a_{N_c-1}) ,
\hskip10mm \s_{1}
=\left( \begin{array}{cc} 0 &  1 \\ 1 & 0 \end{array} \right ).
\en
The scale matching relation becomes
\eq
\La^{2 \cdot (N_c+1)}={\La_L}^{3 \cdot 2 \cdot \frac{1}{2}} 
a_1^4 
\left ( \prod_{i=2}^{N_c-1} (a_1^2-a_i^2) \right )^2 
 (M_{\rm ad})^{-1}.
\en
Since the $SU(2)$ adjoint mass is given by 
$M_{\rm ad}=g_{2 N_c} a_1^2 \prod^{N_c-1}_{i=2} (a_1^2-a_i^2)$ we get
$ {\La_L}^3 = g_{2 N_c} \La^{2(N_c+1)}/a_1^2$. The low-energy
effective superpotential thus turns out to be
\eq
W_L=W_{cl}+2  \frac{g_{2 N_c}}{a_1^2} \La^{2(N_c+1)}.
\en

Checking the result with $Sp(4)$ we encounter the same problem as in the
$SO(5)$ theory. Instead of (\ref{sptree}), thus, we take a superpotential
in the form (\ref{ws2}), reproducing the $N=2$ $Sp(4)$ curve \cite{ArSh}.
Similarly, for $Sp(2 N_c)$ we study a superpotential 
(\ref{ws}). It turns out that $\bra s_i\ket$ are calculated as
\eqn
\langle s_n \rangle & = &  s_n^{cl}(g), \hskip12mm 1 \leq n \leq N_c-2, \CR 
\langle s_{N_c-1} \rangle & = &  s_{N_c-1}^{cl}(g)
- \frac{2}{a_1^4} \La^{2(N_c+1)}, \CR
\langle s_{N_c} \rangle & = &  s_{N_c}^{cl}(g)
+  \frac{4}{a_1^2} \La^{2(N_c+1)}.
\enn
These satisfy the $N=2\; Sp(2 N_c)$ singularity condition \cite{ArSh} 
since the curve exhibits the quadratic degeneracy
\eqn
x^2 y^2 & = & \left( x^2 \left \langle {\rm det} (x-\Phi) 
\right \rangle +\La^{2(N_c+1)} \right)^2
- \La^{4(N_c+1)}  \CR
& = & (x^{2 (N_c+1)} - \langle s_1 \rangle x^{2 N_c} -\cdots 
 - \langle s_{N_c-1} \rangle x^4- \langle s_{N_c} \rangle x^2
+ 2  \La^{2 (N_c+1)} ) \CR
&  & \; \times (x^{2 (N_c+1)} - \langle s_1 \rangle x^{2 N_c} -
\cdots 
 - \langle s_{N_c-1} \rangle x^4- \langle s_{N_c} \rangle x^2) \CR
& = & \left\{ x^2 (x^2-a_1^2)^2 (x^2-a_2^2) \cdots (x^2-a_{N_c-1}
^2)+
2 \La^{2 (N_c+1)} 
\left( \left( \frac{x}{a_1} \right)^4 -
2 \left( \frac{x}{a_1} \right)^2 +1 \right) \right\} \CR
&  & \; \times
\left( x^2 {\rm det} (x-\Phi_{cl}) \right)\CR
& = & (x^2-a_1^2)^2 
\left( x^2 (x^2-a_2^2) \cdots (x^2-a_{N_c-1}^2) +
 \frac{\La^{2( N_c+1)}}{a_1^4}
\right) \times
\left( x^2 {\rm det} (x-\Phi_{cl}) \right).
\enn
It should be mentioned that our remarks on $SO(2 N_c+1)$ theories also
apply here.

%%%%%%%%%%%%%%%%%
%\section{non-trivial $N=1$ SCFT}

Now that we have found the $N=1$ effective superpotentials which can be used
to derive the $N=2$ curves we wish to discuss $N=1$ SCFT.
Recently large classes of novel $N=2$ SCFT have been
shown to exist by fine-tuning the 
moduli coordinates at the points of coexisting mutually non-local
massless dyons in the $N=2$ Coulomb phase 
\cite{ArDo}, \cite{ArPlSeWi}, \cite{EgHoItYa}.
On the other hand, an advantage of the present ``integrating in'' approach
lies in the fact that we can explicitly read off 
how the microscopic parameters in the $N=1$ theory
are related to the $N=2$ moduli coordinates.
Therefore the coupling constants in our effective superpotentials 
are easily adjusted to the values of $N=2$ non-trivial 
critical points. Upon doing so we expect new classes of $N=1$ 
SCFT to be realized. This was first exploited by Argyres and Douglas 
in the $N=2$ $SU(3)$ gauge theory \cite{ArDo}. In the following we shall show 
that a mass gap in the $N=1$ confining phase of $SU(N_c)$ and $SO(2N_c)$
theories vanishes when the $N=1$ parameters are tuned as described above.
Thus non-trivial $N=1$ fixed points will be identified.

We first consider the $N=2$ $SU(N_c)$ theory with $N_c \ge 3$. 
It was shown that the $N=2$
highest critical points exist at $\bra u_i\ket =0$ for $2 \le i\le N_c-1$
and $\bra u_{N_c}\ket =\pm 2\Lambda^{N_c}$ \cite{EgHoItYa}. The critical
points are featured by $Z_{N_c}$ symmetry. When we approach these points
under the $N=1$ perturbation it is clear from (\ref{suneigen}) 
and (\ref{su}) that the coupling constants in (\ref{r1}) become
\eq
g_i  \longrightarrow 0,  \hskip10mm  2 \le i \le N_c-1
\label{znpoint}
\en
and the superpotential reduces to
\eq
W_{\rm crit}= {g_{N_c} \over N_c} \, {\rm Tr}\, \Phi^{N_c}.
\label{Wcrit}
\en

In our $N=1$ theory there exists a mass gap due to dyon condensation and the
gauge field gets a mass by the magnetic Higgs mechanism.
Let us check how the gap behaves in the limit (\ref{znpoint}). For this purpose
it is convenient to consider a macroscopic $N=1$ superpotential $W_m$
obtained from the effective low-energy $N=2$ Abelian theory.
We denote by $A_i$ the $N=1$ chiral superfield of $(N_c-1)$ $N=2$ $U(1)$
multiplets and by $M_i, \tilde M_i$ the $N=1$ chiral superfields of $N=2$
dyon hypermultiplets. The superpotential $W_m$ then takes the form 
\cite{SeWi}, \cite{ArDo}
\eq
W_m=\sqrt{2} \sum_{i=1}^{N_c-1}A_iM_i\tilde M_i +\sum_{i=2}^{N_c} g_iU_i,
\en
where $U_i$ represent the superfields corresponding to ${\rm Tr}\, \Phi^i$
and their lowest components have the expectation values $\bra u_i\ket$.
The equation of motion is given by
\eq
-{g_n \over \sqrt{2}}=\sum_{i=1}^{N_c-1}{\pa a_i \over \pa \bra u_n\ket}
m_i\tilde m_i, \hskip10mm  2 \le n \le N_c
\label{eqmotion1}
\en
and
\eq
a_im_i=a_i\tilde m_i =0, \hskip10mm 1 \le i \le N_c-1
\label{eqmotion2}
\en
where $a_i, m_i$ and $\tilde m_i$ stand for the expectation values of the
lowest components of $A_i, M_i$ and $\tilde M_i$ respectively. The $D$-flatness
condition implies $|m_i|=|\tilde m_i|$.

Concentrate now on the singular point where we have only one massless
dyon, say $M_1, \tilde M_1$. Note that this is the $N=1$ vacuum for which
we have derived the effective superpotential (\ref{wl}). 
Then, $a_1=0$ and $a_i \not= 0$ for $i\not= 1$,
and hence (\ref{eqmotion2}) yields $m_i=0$ for $i\not= 1$.
Eq.(\ref{eqmotion1}) is rewritten as
\eq
{g_i \over g_{N_c}}=
{\pa a_1 /\pa \bra u_i\ket \over \pa a_1 /\pa \bra u_{N_c}\ket},
\hskip10mm  2 \le i \le N_c-1.
\en
We bring the system to the $Z_{N_c}$ critical points by tuning a parameter 
$\epsilon$
\eq
\bra u_n\ket \pm 2\Lambda^{N_c} \delta_{n,N_c}=c_n \epsilon^n,
\hskip10mm c_n={\rm const.}
\en
where $\epsilon$ is an overall mass scale. 
Following \cite{ArDo}, \cite{EgHoItYa} we evaluate
\eq
{\pa  a_1 \over \pa \bra u_i\ket} \simeq \epsilon^{N_c/2+1-i},
\hskip10mm 2 \le i \le N_c.
\en
Thus we find 
\eq
{g_i \over g_{N_c}} \simeq \epsilon^{N_c-i} \longrightarrow 0
\en
for $2 \le i \le N_c-1$. This agrees with (\ref{znpoint}). 

The gap in the $U(1)$ factor arises from the dyon condensation $m_1$. From 
(\ref{eqmotion1}) we obtain the scaling behavior
\eq
m_1 
= \Big( -{g_{N_c}\over \sqrt{2} \pa a_1 /\pa \bra u_{N_c}\ket} \Big)^{1/2}
\simeq  \sqrt{g_{N_c}} \, \epsilon^{(N_c-2)/4} \longrightarrow 0. 
\en
Thus the gap vanishes as we approach the $Z_{N_c}$ critical point.
Therefore the $Z_{N_c}$ vacua of our $N=1$ theory characterized by a 
superpotential (\ref{Wcrit}) is a non-trivial fixed point.

We now turn to the $SO(2N_c)$ theory with $N_c \ge 3$. 
The $N=2$ $SO(2N_c)$ theory possesses 
the highest critical points at $\bra u_{2i}\ket =0$ $(1\le i \le N_c-2),$ 
$\bra v \ket =0$ and $\bra u_{2(N_c-1)}\ket =\pm 2 \Lambda^{2(N_c-1)}$ 
\cite{EgHoItYa}. In the $N=1$ superpotential (\ref{soeventree}) this 
criticality corresponds to
\eq
g_{2i} \longrightarrow 0, \hskip10mm   \lm \longrightarrow 0
\label{socrit}
\en
for $1 \le i \le N_c-2$ and we have
\eq
W_{\rm crit}= {g_{2(N_c-1)} \over 2(N_c-1)} \, {\rm Tr}\, \Phi^{2(N_c-1)}.
\label{soWcrit}
\en

Let us show that an $N=1$ gap vanishes in this limit by looking at the 
singular point where a single massless dyon exists. A similar analysis to
the $SU(N_c)$ theory gives us the vacuum condition
\eqn
{g_{2i} \over g_{2(N_c-1)}}=
{\pa a_1 /\pa \bra u_{2i}\ket \over \pa a_1 /\pa \bra u_{2(N_c-1)}\ket},
\hskip10mm 
{\lm \over g_{2(N_c-1)}}=
{\pa a_1 /\pa \bra v \ket \over \pa a_1 /\pa \bra u_{2(N_c-1)}\ket}
\label{vaqcon}
\enn
for $1 \le i \le N_c-2$.
The critical limit is taken through the parametrization
\eqn
&& \bra u_{2n}\ket \pm 2\Lambda^{2(N_c-1)}\delta_{n, N_c-1}=c_n \epsilon^{2n},
\hskip10mm  1 \le n \le N_c-1,  \CR
&& \bra v \ket=c\, \epsilon^{N_c},
\enn
where $\epsilon$ is an overall mass scale and $c_n, c$ are 
$\epsilon$-independent constants. We obtain from (\ref{vaqcon}) that
\eqn
&& {g_{2i} \over g_{2(N_c-1)}} \simeq \epsilon^{2(N_c-1-i)} \longrightarrow 0,
\hskip10mm   1 \le i \le N_c-2, \CR
&& {\lm \over g_{2(N_c-1)}} \simeq \epsilon^{N_c-2} \longrightarrow 0
\enn
in agreement with (\ref{socrit}). The gap in the $U(1)$ factor scales as
\eq
m_1 
= \Big( -{g_{2(N_c-1)}
\over \sqrt{2} \pa a_1 /\pa \bra u_{2(N_c-1)}\ket} \Big)^{1/2}
\simeq  \sqrt{g_{2(N_c-1)}} \, \epsilon^{(N_c-2)/2} \longrightarrow 0. 
\en
Thus our $N=1$ $SO(2N_c)$ theory with a superpotential (\ref{soWcrit}) has a 
non-trivial fixed point.

In conclusion we have shown how to derive the curves for the Coulomb phase 
of $N=2$ Yang-Mills theories with classical gauge groups 
by means of the $N=1$ confining phase superpotential.
Transferring the critical points in the $N=2$ Coulomb phase 
to the $N=1$ theories we have found non-trivial $N=1$ SCFT with 
the adjoint matter governed by a superpotential.
This SCFT certainly has a connection with the non-Abelian
Coulomb phase of the Kutasov-Schwimmer model 
\cite{Ku}, \cite{KuSc}, \cite{KuScSe}.
To further explore this connection it will be interesting to
investigate theories containing the additional fundamental matter multiplets.

%%%%%%%%%%%%%%%%%%%%%%%
%\section{Acknowledgements}

\vskip10mm
S.K.Y. would like to thank T. Eguchi for discussions on 
$N=1$ SCFT with a superpotential.
The work of S.T. is supported by JSPS Research Fellowship for Young 
Scientists. The work of S.K.Y. is supported in part by the
Grant-in-Aid for Scientific Research on Priority Area 213 
``Infinite Analysis'', the Ministry of Education, Science and Culture, Japan.

\newpage

%%%%%%%  References

\end{document}